\documentclass[10pt]{article}
\usepackage{amssymb}
\usepackage{amsthm}
\usepackage{authblk}
\usepackage{setspace}
\usepackage[textwidth=6.5in,textheight=8.25in,centering]{geometry}
\usepackage[pdftex]{graphicx}

\onehalfspace
\begin{document}
\begin{center}
{\LARGE{\textbf{Malignancy Induced Subtle Perturbation Sensitive Raman Scattering for Glioma Detection and Grading}}}

\vspace{0.5 cm}
\textit{Ritika Kaushik$^{\$ , 1}$, Chanchal Rani $^{ \$ , 1}$,  Km.  Neeshu  $^{ \$ , 1}$, Manushree Tanwar,  $^{ \$ , 1}$, Devesh K. Pathak$^{ \$ , 1}$, Anjali Chaudhary$^{1}$, Fouzia Siraj$^{2}$, Hem C. Jha,$^{ 3}$ and Rajesh Kumar$^{* ,1}$}
\vspace{0.5 cm}

$^{1}$ Materials and Device laboratory, Discipline of Physics, Indian Institute of Technology Indore, Simrol-453552, India

$^{2}$ National Institute of Pathology, (Indian Council of Medical Research), Safdarjung Hospital Campus, New Delhi. India. 110029

$^{3}$ Discipline of Biosciences and Biomedical Engineering, Indian Institute of Technology Indore, Simrol, India-453552

$^{\$} $ Authors contributed equally

$^{*} $ Corresponding Author: rajeshkumar@iiti.ac.in

\vspace{1 cm}
ABSTRACT

\end{center}
Subtle changes in Raman spectral line-shape have been observed from malignant human brain cells and its possibility for being used in detection and grading of Glioma has been explored here. The latter has been developed as a result of the fact that the width of the Raman spectra is more sensitive, as compared to the peak position, to the brain tumors. The perturbations induced by the cell-modification, as a consequence to the cancerous growth, may be responsible for the width’s variation in the Raman spectrum due to vibrational lifetime alteration enforced at the molecular levels. A consistent cancer induced effect on the spectral width has been observed for three different brain cells’ Raman modes at frequencies 1001 cm$^{-1}$, 1349 cm$^{-1}$ and 1379 cm$^{-1}$. Raman spectral analysis reveals that for cancerous cells, the FWHM varies up to 35 \% in comparison with the healthy cells. It has been established how a careful analysis of Raman spectra can help in easy detection of brain tumors. The methodology has been validated by studying the effect of similar microscopic perturbations, e.g, Fano coupling and quantum size effects, on different Raman spectral parameters which also reveals Raman width to be the most sensitive parameter.      
\vspace{0.5cm}


\vspace{0.2cm}

\newpage
\section{Introduction}
Raman scattering[1,2], since its discovery in 1928, has now been established as a spectroscopic technique[3–-6] having many applications in different fields of science, engineering and technology[7,8]. Beyond identification of bond modes, a Raman spectrum is capable of investigating several materials’ properties if analysed under appropriate framework. Mode identification can be done by indexing the Raman peak positions but it may contain lot more information than simple chemical bonds energies which may also be investigated using IR spectroscopy, of course with certain limitations. Beyond the peak position of a Raman peak, the overall line-shape of the spectrum is equally important as it may reveal several important physical phenomena taking place at microscopic level like quantum confinement or size effect, electron-phonon effect, fantum effect etc[9–-16].  Apart from direct evidences, as a consequence of Raman line-shape’s sensitivity towards external conditions, perturbations induced by temperature, pressure etc also can be investigated using classical and advanced Raman techniques[17–-21]. The external perturbations like temperature, pressure etc. induce the Raman line-shape to alter by means of change in its peak position, width, asymmetry etc as compared to unperturbed condition. These perturbation induced line-shape changes have allowed Raman spectroscopy to be used in bioscience as a diagnostic tool. In biosciences, Raman spectroscopy also bears certain advantages over IR spectroscopy due to involvement of water molecules in cells which do not hinder Raman scattering unlike IR spectroscopy.  Still there is a scope for Raman spectroscopy to be used to its full potential through careful analysis of subtler changes in the Raman spectra from biological samples suffering from infections/malignancies[3,22–-28]. 

According to a report of the World Health Organisation, cancer is the second leading cause of death globally and is responsible for an estimated 9.6 M deaths in 2018 with about 1 in 6 deaths take place due to cancer[29]. As known, cancer is the transformation of normal cells into cancerous cells in a multistage process that generally progresses from a pre-cancerous lesion to a malignant tumour. Gliomas are the most common primary brain tumors characterized by diffuse infiltration and aggressiveness.. The presence of an isocitrate dehydrogenase (IDH) mutation was first discovered in colorectal cancers. Parsons et al. [30] found mutations of the IDH1 in 12 \% of the glioblastomas (GBMs)[31–34]. Other large scale studies have also validated that IDH1 and IDH2 mutations were found in the majority of secondary GBM and lower grade gliomas like diffuse astrocytomas (DA), while these were rarely found in adult primary and pediatric GBMs. Looking at the cancer related fatalities, it is important to diagnose the disease early so that necessary therapeutic interventions can be initiated for optimum patient care. In the last couple of decades, classic and advanced Raman spectroscopic techniques like surface enhanced and Fourier transform Raman spectroscopy have been used widely with the latter having certain advantages[32–38]. Raman spectroscopy enables one to retrieve a molecular signature of tissue’s biochemical composition in order to identify tumor and normal tissue, and in conjunction with different statistic algorithms, the spectral data with various pathologic attributions can be differentiated and classified depending on their spectral differences, e.g., peak area, peak height or peak- or spectral line shape. This makes Raman spectroscopy a very efficient analytical technique for rapid and non-destructive diagnosis of human diseases like cancer. One of the drawbacks in using Raman spectroscopy lies with the tedious spectral analysis involved. This limits the technique’s applicability. A simpler methodology can be developed for the analysis of Raman spectra for easier application in detection of disease which can be done as follows. Disease, an infection or malignancies, can be considered as a transformation of a healthy cell into a deformed state due to an external invasion or unwanted change at the cellular levels. The extent up to which the invasion or malignancy has taken place will affect the healthy cell accordingly. Since the cells are a combination of molecular bonds, any change in terms of infection/malignancy will certainly perturb the molecular structure and thus is expected to manifest itself in the corresponding Raman spectrum. The modification in the Raman spectra, by means of change in its peak position, width, asymmetry etc, thus can be used to identify the nature and quantum of invasion/malignancy after simpler calibrations. This can be used for identification of disease by simple analysis of Raman spectra without going into tedious methodology like principle component analysis. 

The aim of the present article is to suggest a methodology to develop a simpler analytical approach to use Raman spectroscopy as a diagnostic tool for identification and grading of brain cancer. For this purpose, Raman spectra from known samples of malignant human brain cells have been recorded and compared with corresponding normal (healthy) counterpart for analysis of the difference between the two sets of samples. The known cases of gliomas have been established using traditional biopsy samples with histopathological and immonohistochemical analysis..  From the samples, on which the disease has been confirmed, only subtle changes in Raman spectral width has been observed and found sufficient to establish and grade the presence of cancer. The methodology has been validated by studying a perturbation induced Raman spectral changes in a known silicon (Si) nanowires’ systems. Three different types of perturbations in terms of Fano coupling and quantum confinement effects have been investigated to reveal that Raman spectral width remains the most sensitive parameter to get affected by subtle perturbations. This is similar to the biological observations and thus helps establishing the proposed methodology for disease diagnosis and grading.

\section{Experimental Details}

Samples comprised biopsies of patients presenting with intracranial mass lesions to Safdarjung Hospital, New Delhi, India. Histopathological examination of biopsy material was performed at National Institute of Pathology, New Delhi. Hematoxylin and eosin staining was done followed by immunohistochemistry for IDH 1 mutation. Optical microscope (Nikon) was used for histopathological analysis. The morphological images of the samples were observed by fluorescence inverted  microscope (Nikon$^{TM}$ make) under ambient condition using Differential Interference Contrast (DIC) for two malignant samples as well as the healthy sample used as control. The three samples included normal brain, Diffuse Astrocytoma IDH-mutant, Grade II and Glioblastoma IDH- wild type, Grade IV. Grading was done according to WHO Classification of Central Nervous system tumors (2016).  Raman spectra from all of these samples have been recorded using Horiba-JobinYvon (Labram-HR) spectrometer using 633 nm excitation source in back scattering geometry.  The Raman width analyses have been done by carrying out a Gaussian fitting of the Raman spectra from which the experimental widths have been estimated for rest of the study.

\section{Results and Discussion}
For diagnosis and grading of tumors, the biopsy samples were subjected to histopathological examination. The H \& E stained slides were viewed under an optical microscope (Figure 1a, 1b) and the same were compared with the normal brain parenchyma (Figure 1c). The IDH mutation analysis was done by immunohistochemistry using Anti IDH-1 antibody (Dianova). The samples were classified as Diffuse astrocytoma, IDH- mutant (DA) (Figure 1a), Glioblastoma, IDH- wildtype (GBM) (Figure 1b) and normal brain (Figure 1c). These terminologies DA, IDH mutant, GBM and Normal will be used for Diffuse Astrocytoma, Isocitrate Dehydrogenase mutant, Glioblastoma  IDH wildtype and normal brain parenchyma respectively throughout the manuscript.

\begin{figure}
\begin{center}
\includegraphics[width=16cm]{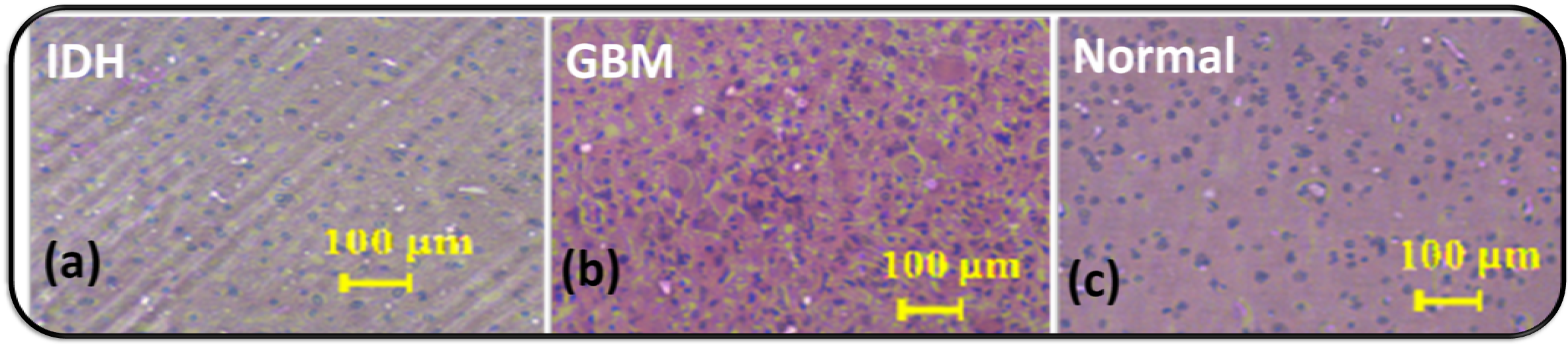}
\caption{Optical microscopic images from (a) Diffuse astrocytoma  (IDH-mutant) (WHO Grade II), (b) Glioblastoma (GBM, IDH wildtype) (WHO Grade IV) and (c) of the normal brain tissue sample (no tumor present).}
\end{center}
\end{figure}

\begin{figure}
\begin{center}
\includegraphics[width=15cm]{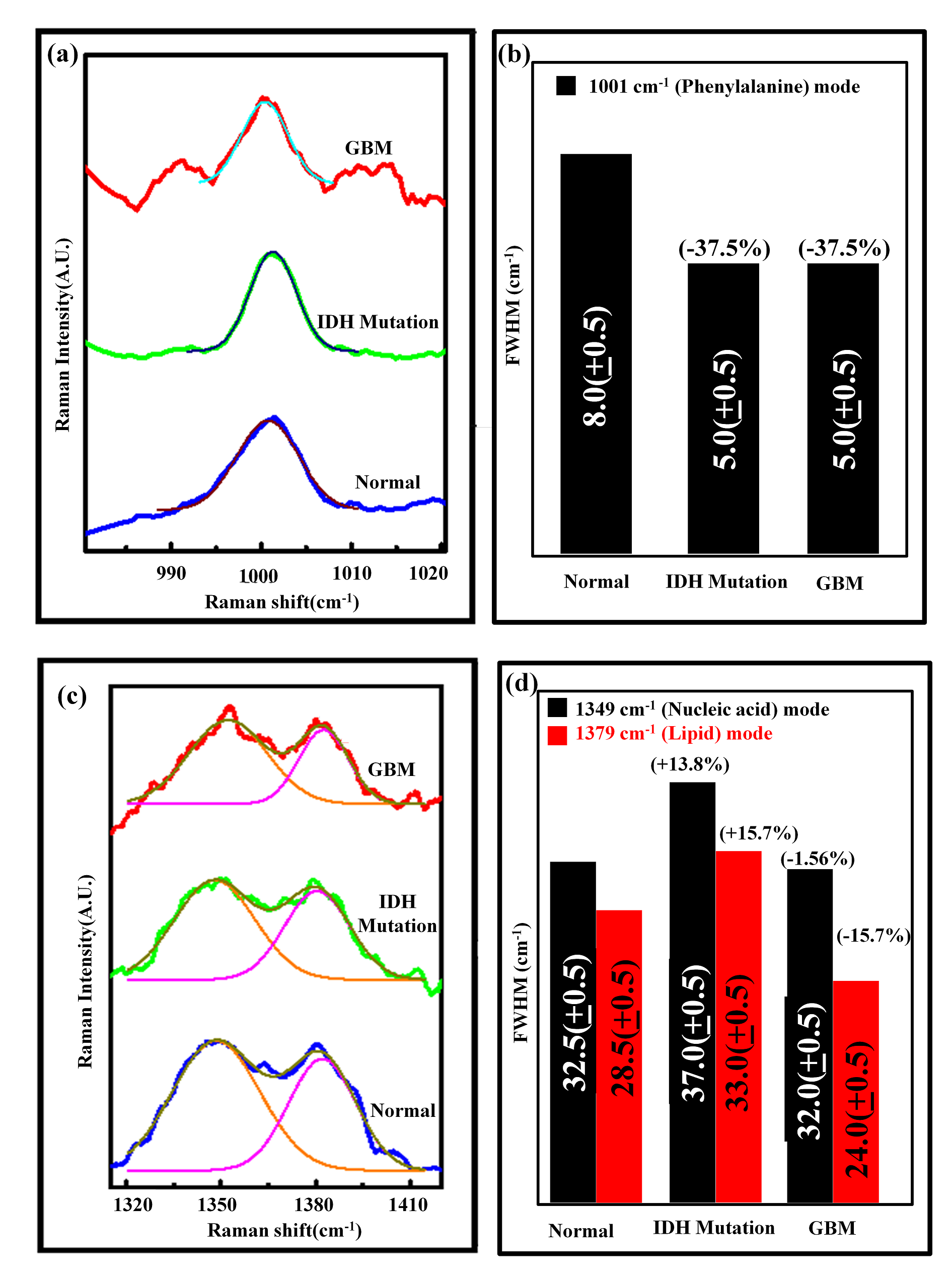}
\caption{Raman spectra from different samples showing (a) the 1001 cm$^{-1}$ phenylalanine mode and comparison of the observed FWHM values (b); (c) the nucleic acid, 1349 cm$^{-1}$ and lipid, 1379 cm$^{-1}$ modes along with the comparative FWHM variation (d). }
\end{center}
\end{figure} 

Observation of distinct morphologies from different grade cells (Figure 1) clearly suggests different microstructures at the microscopic level. Since these cells are composed of different molecules thus subtle changes must have taken place at micro-level. If these changes can sufficiently perturb the physical nature of the cells, which is actually likely from Figure 1, the same must be observable in Raman spectra. Keeping the above hypothesis in mind, Raman spectra from these samples have been recorded (Figure 2). Raman spectra from all the samples exhibited three modes around $\sim$1001 cm $^{-1}$ (mode of protein) , 1349 cm $^{-1}$ (mode of nucleic acid) and 1379 cm $^{-1}$ (mode of lipids) which is consistent with the literature [23] and have been identified for detailed Raman analyses. The Raman peak position of all these modes remain unchanged even from the malignant cells. This immunity of Raman peak position with respect to the presence of disease means that the disease induced perturbations are not sufficient to change the frequency of the corresponding molecular vibrations and are rather subtle in nature if at all present. A closer analysis reveals that the full width at half maximum (FWHM) of the protein mode (1001 cm$^{-1}$) varied significantly by 37\% (decrease) for DA, IDH mutant as well as GBM stages of the disease, as compared to the normal samples for which the Raman mode was 8 cm$^{-1}$ wide that reduces to 5 cm$^{-1}$ for the disease stages (Figure 2b). Similar trend was also observed for the nucleic acid and lipid modes (Figure 2c) with non varying frequencies at 1349 cm $^{-1}$ and 1379 cm $^{-1}$ respectively. Trend in disease dependent width variation was observed for these two modes (Figure 2d) showing $\sim$13.8 \% change (increase) in mode width for DA IDH mutant and $\sim$2\% change (decrease) in mode width for GBM case as compared to normal cells where the width was 32.5 cm$^{-1}$ for the 1349 cm$^{-1}$ mode (black bars, Figure 2d). On the other hand, the 1379 cm$^{-1}$ mode showed an increase (decrease) of 15.7\% in FWHM for IDH (GBM) disease stages as compared to the 28.5 cm$^{-1}$ wide Raman mode (red bars, Figure 2d) from healthy cells. The results were found consistent, in terms of unchanged peaks position and varying FWHM, for all the three modes’ analyses done with a sample size of 25. The obtained results, the disease dependent variation in only the FWHM with mode frequency remaining unaffected, are unprecedented. Such a disease dependent observation in Raman width can be useful in identifying and grading the disease, due to consistent results, but needs validation and explanation of such observation. In other words, if significant and consistent changes in FWHM are observed, it can be assigned due to malignancy if other infections have been ruled out.
 
 As mentioned above, the sole variation of the Raman mode, with little effect in the peak position, clearly indicated that the disease dependence perturbation is subtle in nature i.e, it is not sufficient to change the frequency of the corresponding molecular vibration. The exclusive variation in the width could be likely due to the malignant induced change in the vibrational life time of the cancerous cell. This has been validated by taking an example of Raman line-shape variation as a result of systematic perturbation in a low dimensional solid state semiconductor system. Raman line-shape analysis has been done from a very well known semiconductor system, Si, where quantum confinement effect and electron-phonon (Fano) interaction have been used as known control perturbations and will be discussed one-by-one for validation.  For this purpose, the theoretical Raman line-shapes have been generated using the universal perturbation dependent Raman line-shape function developed earlier by Richter et al[39] and later modified by Campbel et al[40,41] and several others[5,6,10,15,42–-46]. The simplified general equation for universal Raman line-shape can be represented by Eq. 1 below: 
 
 \begin{equation}
I(\omega) = \int _0 ^1 e^{-\frac{k^2L^2}{4a^2}}\left[ \frac{(\epsilon +q)^2}{1+\epsilon ^2}\right ]d^nk
\end{equation}

  where,  $\epsilon = \frac{\omega - \omega(k)}{\gamma /2}$, D, a, $\gamma$ denote the nanocrystallite size, lattice constant and line width (~$\sim$ 4 cm$^{-1}$) of Si respectively. Here ‘n=2’ is the degree of quantum confinement for the case of SiNWs. The  $\omega ^2(k)$ = 171400 + 100000 cos ($\pi$k/2))   is the phonon dispersion relation for Si and `q’ is the Fano asymmetry parameter which is the measure of extent of Fano interaction present in the system. To see the effect of Fano coupling on line-shape variation, the line-shape generated using different `q’ values between -3 to -10 have been analyzed (Figure 3a) representing a donor-type discrete-continuum Fano interference [14,43–45,47]12,38–40,42 for a given crystallite size of 3 nm. The obtained theoretical line-shapes are broad and asymmetric (as compared to crystalline Si counterpart which is sharp and symmetric[4,6,46] in nature as a consequence of the classic discrete-continuum (Fano-) interaction[48]44. Figure 3a clearly shows that the Raman line-shape becomes more and more asymmetry when the absolute value of q reduces (due to increase in Fano coupling). The line-shape asymmetry is quantified using the term asymmetry ratio ($\alpha$) defined by the ration $\alpha$ = $\gamma _l$/ $\gamma _h$, where $\gamma _l$ and $\gamma _h$ are the spectral half widths of the line-shape towards the lower and higher energy side of the Raman peak position respectively. It is apparent that with changing q values, the width, $\gamma$ (= $\gamma _l$ + $\gamma _h$) as well as the peak position ($\omega$) both changes but not at the same rate (with respect to ‘q’). The rate of change of γ and ω has been plotted as a function of q in Figure 3b which shows that the FWHM changes more rapidly as compared to the peak position (i.e, $d \gamma / dq > d \omega / dq $). In physical terms, it means that the Raman spectral width is more sensitive to the micro level perturbation induced by donor type Fano coupling. This behavior is similar to the one discussed above (Figure 2) where Raman spectral width exhibits change without affecting the peak position which is expected to change only after the perturbation is of sufficient quantum. This is also apparent from Figure 3b where the peak position does not change until a large Fano coupling is present whereas the width starts responding to the perturbation relatively earlier.
\begin{figure}
\begin{center}
\includegraphics[width=12cm]{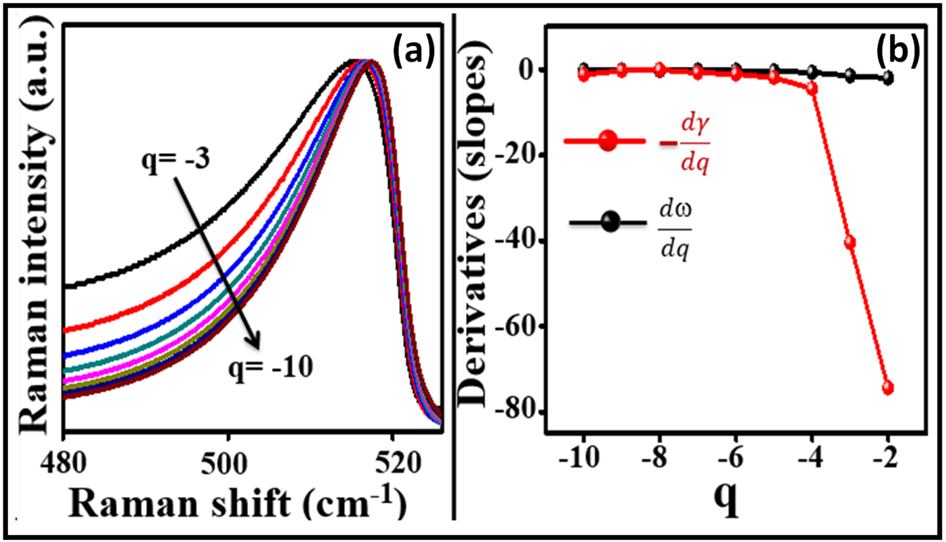}
\caption{(a) Theoretical Raman line shapes generated using Eq. 1 for different values of Fano coupling parameter q and crystallite size D = 3 nm, (b) Rate of change of variation of width and Raman peak shift as a function of Fano parameter showing width being the fastest varying parameter for donor type Fano coupling system}
\end{center}
\end{figure}

\begin{figure}
\begin{center}
\includegraphics[width=12cm]{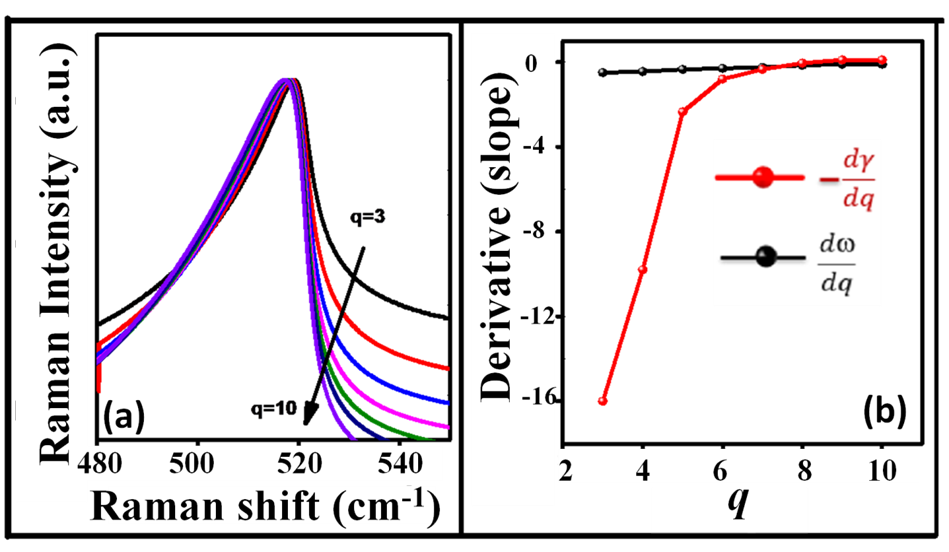}
\caption{(a) Theoretical Raman line shapes generated using Eq. 1 for different values of Fano coupling parameter q and crystallite size D = 3 nm, (b) Rate of change of variation of width and Raman peak shift as a function of Fano parameter showing width being the fastest varying parameter for acceptor type Fano coupling system. }
\end{center}
\end{figure} 

\begin{figure}
\begin{center}
\includegraphics[width=12cm]{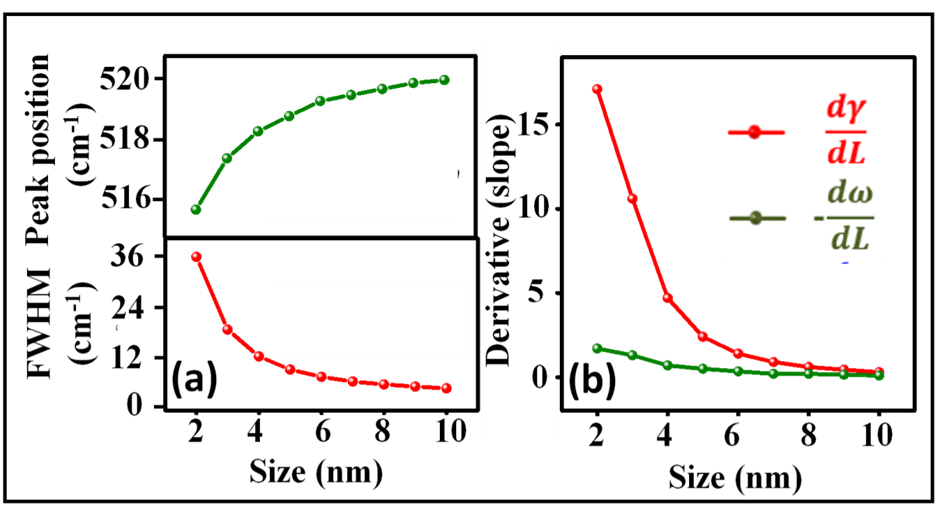}
\caption{(a) Variation of Raman peak position and width as a function of crystallite size, obtained using Eq. 1 using phono confinement model and (b) Rate of change of variation of width and Raman peak shift as a function of size showing width being the fastest varying parameter for systems with quantum confinement effect.}
\end{center}
\end{figure} 

Before concluding the above-mentioned observation, the Fano parameter dependent Raman spectral line-shape variation has been studied for an acceptor type discrete-continuum Fano coupling as shown in Figure 4a and corresponding rate of change of line-shape parameters has been shown in Figure 4b.  However, this system is known to be complementary to the case discussed in Figure 3, the sensitivity of the width remains prominent as compared to the peak position or asymmetry ratio here too. It further endorses the conclusions drawn above that Raman spectral line width ($\gamma$) remains the most sensitive parameter which starts responding to even subtle perturbations whereas stronger perturbations are needed to be reflected in other Raman line-shape features like peak position. This further affirms that the disease induced width variation can potentially be used to identify presence of malignancy in cells. Furthermore, to validate the above-said fact (the FWHM’s sensitivity to subtle perturbation) another perturbation, of completely different in nature, the quantum confinement (or size-) effect, has also been considered as described somewhere else in great detail[49].  For this analysis, Raman line-shapes obtained corresponding to different Si nanowires’ size have been analyzed[49] with no Fano effect. With decrease in size, effect on Raman peak position and FWHM becomes prominent (Figure 5a) with the latter varying more rapidly as evident from the derivative plot in Figure 5b.

Three different perturbations, donor type Fano interference (Figure 3), acceptor type Fano interference (Figure 4) and quantum confinement effect (Figure 5) [49] reveal consistently that all these microscopic level perturbations can affect the Raman spectral widths even with their minimum possible presence. On the other hand, stronger perturbations are necessary to induce changes in the Raman peak position.   In other words, a change in Raman spectral width ($\gamma$) with little associated changes in other Raman parameters, in a system can be observed due to subtle changes at microscopic level in a system and the same should not be neglected as it may contain important information. Since the malignancy can be considered analogous to these perturbations, thus the malignancy induced modifications observed in the spectral widths from human brain cells are sufficient to assign the changes due to the developement of glioma in the patient and thus in fact can be used as a diagnostic tool for identification and grading of the disease. It also opens up a possibility to explore the use of this simple observation as a tool for early detection of cancer without going into tedious Raman spectroscopic analysis.

\section {Conclusions}
Raman spectroscopic studies from human brain cells, suffering from Glioma, duly diagnosed using traditional methods from the biopsy samples, shows variations in spectral widths with little effect on the Raman peak positions as compared with the healthy cells. The above mentioned variations have been observed as a result of malignancy induced perturbations at the molecular level. The protein, lipid and nucleic acid Raman vibrational modes’ width change by upto $\sim$38\% from the malignant cells as compared to the healthy cells. Above observations can be used for identification and grading of the disease in a simpler manner as the results have been validated using a nanostructured system where these subtle perturbations are  observed to be inducing changes in the width as it is the most sensitive parameter to get affected by minimal perturbations. Effect of donor- and acceptor- type continuum-discrete interaction and quantum dimensional size on Raman spectral parameters confirms that spectral width is the most sensitive parameter to any subtle perturbations. This helps in deducing and validating the fact that the malignancy actually acts as a physical cause to induce enough perturbation to affect the phonon life time of the molecular level to be reflected in the width and thus can be used for disease detection and staging using the simpler Raman spectroscopy based technique.

\section*{Acknowledgements} 
Authors acknowledge Sophisticated Instrumentation Centre (SIC), IIT Indore for SEM measurements. Authors thank funding received from Science and Engineering Research Board (SERB), Govt. of India (grant no. CRG/2019/000371). Authors thank IIT Indore for providing fellowship. Author (D.K.P.) acknowledges Council of Scientific and Industrial Research (CSIR) for financial support (file number 09/1022(0039)/2017-EMR-I) and author (M.T.) thank Department of Science \& Technology (DST), Govt. of India, for fellowship (file no. DST/INSPIRE/03/2018/000910/IF180398). Facilities received from Department of Science and Technology (DST), Govt. of India, under FIST Scheme with grant number SR/FST/PSI-225/2016 is also acknowledged. Thanks to Dr. P.R. Sagdeo (IIT Indore, India), Dr. S.K. Saxena (University of Alberta, Canada) and Dr. P Yogi (LUH Hannover, Germany) for useful discussions.

\textbf{Authors declare no conflicts of interest.}

\newpage


\begin{thebibliography}{49}

\bibitem{1.}	C. Raman, Indian J. Phys. 1928, 02, 387.
\bibitem{2.}		C. V. Raman, K. S. Krishnan, Nature 1928, 121, 501.
\bibitem{3.}		S. Kumar, A. Visvanathan, A. Arivazhagan, V. Santhosh, K. Somasundaram, S. Umapathy, Anal. Chem. 2018, 90, 12067.
\bibitem{4.}		P. Yogi, M. Tanwar, S. K. Saxena, S. Mishra, D. K. Pathak, A. Chaudhary, P. R. Sagdeo, R. Kumar, Anal. Chem. 2018, 90, 8123.
\bibitem{5.}		M. Tanwar, A. Chaudhary, D. K. Pathak, P. Yogi, S. K. Saxena, P. R. Sagdeo, R. Kumar, J. Phys. Chem. A 2019, 123, 3607.
\bibitem{6.}		S. K. Saxena, R. Borah, V. Kumar, H. M. Rai, R. Late, V. G. Sathe, A. Kumar, P. R. Sagdeo, R. Kumar, Journal of Raman Spectroscopy 2016, 47, 283.
\bibitem{7.}		A. Chaudhary, D. K. Pathak, M. Tanwar, R. Kumar, Anal. Chem. 2020, 92, 6088.
\bibitem{8.}		R. Kumar, R. G. Pillai, N. Pekas, Y. Wu, R. L. McCreery, J. Am. Chem. Soc. 2012, 134, 14869.
\bibitem{9.}	E. G. Barbagiovanni, D. J. Lockwood, P. J. Simpson, L. V. Goncharova, J. Appl. Phys. 2012, 111, 034307.
\bibitem{10.}		S. K. Saxena, P. Yogi, S. Mishra, H. M. Rai, V. Mishra, M. K. Warshi, S. Roy, P. Mondal, P. R. Sagdeo, R. Kumar, Phys. Chem. Chem. Phys. 2017, 19, 31788.
\bibitem{11.}		K. W. Adu, H. R. Gutiérrez, U. J. Kim, P. C. Eklund, Phys. Rev. B 2006, 73, 155333.
\bibitem{12.}		P. Miska, M. Dossot, T. D. Nguyen, M. Grün, H. Rinnert, M. Vergnat, B. Humbert, J. Phys. Chem. C 2010, 114, 17344.
\bibitem{13.}		N. Fukata, Advanced Materials 2009, 21, 2829.
\bibitem{14.}		P. Yogi, S. Mishra, S. K. Saxena, V. Kumar, R. Kumar, J. Phys. Chem. Lett. 2016, 7, 5291.
\bibitem{15.}		P. Yogi, D. Poonia, S. Mishra, S. K. Saxena, S. Roy, V. Kumar, P. R. Sagdeo, R. Kumar, J. Phys. Chem. C 2017, 121, 5372.
\bibitem{16.}		A. S. Krylov, S. N. Sofronova, I. A. Gudim, S. N. Krylova, R. Kumar, A. N. Vtyurin, J. Adv. Dielect. 2018, 08, 1850011.
\bibitem{17.}		A. Torres, A. Martín-Martín, O. Martínez, A. C. Prieto, V. Hortelano, J. Jiménez, A. Rodríguez, J. Sangrador, T. Rodríguez, Applied Physics Letters 2010, 96, 011904.
\bibitem{18.}		J. Niu, J. Sha, D. Yang, Scripta Materialia 2006, 55, 183.
\bibitem{19.}		B. G. Oscar, C. Chen, W. Liu, L. Zhu, C. Fang, J. Phys. Chem. A 2017, 121, 5428.
\bibitem{20.}		C. Ulrich, E. Anastassakis, K. Syassen, A. Debernardi, M. Cardona, Phys. Rev. Lett. 1997, 78, 1283.
\bibitem{21.}		B. A. Weinstein, G. J. Piermarini, Phys. Rev. B 1975, 12, 1172.
\bibitem{22.}		Z. Movasaghi, S. Rehman, D. I. U. Rehman, Applied Spectroscopy Reviews 2007, 42, 493.
\bibitem{23.}		A. C. S. Talari, Z. Movasaghi, S. Rehman, I. ur Rehman, Applied Spectroscopy Reviews 2015, 50, 46.
\bibitem{24.}		S. Kumar, R. Gopinathan, G. K. Chandra, S. Umapathy, D. K. Saini, Anal Bioanal Chem 2020, 412, 2505.
\bibitem{25.}		R. J. Swain, S. J. Kemp, P. Goldstraw, T. D. Tetley, M. M. Stevens, Biophysical Journal 2008, 95, 5978.
\bibitem{26.}		Q. Tu, C. Chang, Nanomedicine: Nanotechnology, Biology and Medicine 2012, 8, 545.
\bibitem{27.}		K. Kong, C. Kendall, N. Stone, I. Notingher, Advanced Drug Delivery Reviews 2015, 89, 121.
\bibitem{28.}		A. Mahadevan-Jansen, R. Richards-Kortum, in Proceedings of the 19th Annual International Conference of the IEEE Engineering in Medicine and Biology Society. ‘Magnificent Milestones and Emerging Opportunities in Medical Engineering’ (Cat. No.97CH36136), 1997, vol. 6, pp. 2722–2728 vol.6.
\bibitem{29.}		Cancer, https://www.who.int/news-room/fact-sheets/detail/cancer, (accessed 25 April 2020).
\bibitem{30.}		D. W. Parsons, S. Jones, X. Zhang, J. C.-H. Lin, R. J. Leary, P. Angenendt, P. Mankoo, H. Carter, I.-M. Siu, G. L. Gallia, A. Olivi, R. McLendon, B. A. Rasheed, S. Keir, T. Nikolskaya, Y. Nikolsky, D. A. Busam, H. Tekleab, L. A. Diaz, J. Hartigan, D. R. Smith, R. L. Strausberg, S. K. N. Marie, S. M. O. Shinjo, H. Yan, G. J. Riggins, D. D. Bigner, R. Karchin, N. Papadopoulos, G. Parmigiani, B. Vogelstein, V. E. Velculescu, K. W. Kinzler, Science 2008, 321, 1807.
\bibitem{31.}		S. Koljenović, L.-P. Choo-Smith, T. C. Bakker Schut, J. M. Kros, H. J. van den Berge, G. J. Puppels, Laboratory Investigation 2002, 82, 1265.
\bibitem{32.}		C. Krafft, L. Neudert, T. Simat, R. Salzer, Spectrochimica Acta Part A: Molecular and Biomolecular Spectroscopy 2005, 61, 1529.
\bibitem{33.}		A. Mizuno, H. Kitajima, K. Kawauchi, S. Muraishi, Y. Ozaki, Journal of Raman Spectroscopy 1994, 25, 25.
\bibitem{34.}		A. Mahadevan-Jansen, R. R. Richards-Kortum, JBO 1996, 1, 31.
\bibitem{35.}		T. M. Cotton, J.-H. Kim, G. D. Chumanov, Journal of Raman Spectroscopy 1991, 22, 729.
\bibitem{36.}		J. Desroches, M. Jermyn, M. Pinto, F. Picot, M.-A. Tremblay, S. Obaid, E. Marple, K. Urmey, D. Trudel, G. Soulez, M.-C. Guiot, B. C. Wilson, K. Petrecca, F. Leblond, Scientific Reports 2018, 8, 1792.
\bibitem{37.}		T. Hollon, S. Lewis, C. W. Freudiger, X. S. Xie, D. A. Orringer, Neurosurg Focus 2016, 40, E9.
\bibitem{38.}		M. Jermyn, J. Desroches, J. Mercier, K. St-Arnaud, M.-C. Guiot, F. Leblond, K. Petrecca, Biomed. Opt. Express, BOE 2016, 7, 5129.
\bibitem{39.}		H. Richter, Z. P. Wang, L. Ley, Solid State Commun. 1981, 39, 625.
\bibitem{40.}		I. H. Campbell, P. M. Fauchet, Solid State Commun. 1986, 58, 739.
\bibitem{41.}		P. M. Fauchet, I. H. Campbell, Crit. Rev. Solid State 1988, 14, s79.
\bibitem{42.}	J. Zi, K. Zhang, X. Xie, Phys. Rev. B 1997, 55, 9263.
\bibitem{43.}		R. Gupta, Q. Xiong, C. K. Adu, U. J. Kim, P. C. Eklund, Nano Lett. 2003, 3, 627.
\bibitem{44.}		R. Kumar, H. S. Mavi, A. K. Shukla, V. D. Vankar, J. Appl. Phys. 2007, 101, 064315.
\bibitem{45.}		M. Tanwar, D. K. Pathak, A. Chaudhary, P. Yogi, S. K. Saxena, R. Kumar, J. Phys. Chem. C 2020, 124, 6467.
\bibitem{46.}		R. Kumar, G. Sahu, S. K. Saxena, H. M. Rai, P. R. Sagdeo, Silicon 2014, 6, 117.
\bibitem{47.}		F. Cerdeira, M. Cardona, Phys. Rev. B 1972, 5, 1440.
\bibitem{48.}		U. Fano, Phys. Rev. 1961, 124, 1866.
\bibitem{49.}		K. Neeshu, C. Rani, R. Kaushik, M. Tanwar, D. Pathak, A. Chaudhary, A. Kumar, R. Kumar, Advances in Materials and Processing Technologies 2020, 0, 1.





\end{thebibliography}
\end{document}